\documentstyle[psfig,conf_iap]{article}                                    
\begin{document}                                                                
\heading{%
Results from the REFLEX Cluster Survey                                          
}                                                                               
\par\medskip\noindent                                                           
\author{%
H. B\"ohringer$^{1}$, L. Guzzo$^2$, C.A. Collins$^3$, D.M. Neumann$^4$, S. Schindler$^3$,       
P. Schuecker$^{1}$, R. Cruddace$^5$, G. Chincarini$^2$, S. De Grandi$^2$, A.C. Edge$^6$,         
H.T. MacGillivray$^7$, P. Shaver$^8$, G. Vettolani$^9$, W. Voges$^{1}$                      
}                                                                               
\address{                                                                       
Max-Planck-Institut f\"ur extraterr. Physik, D-85740 Garching, Germany          
}                                                                               
\address{%
Osservatorio Astronomico di Brera, Milano/Merate, Italy                                                                
}                                                                               
\address{                                                                     
Liverpool John-Moores University, Liverpool, U.K.,
}                                                                              
\address{                                                                     
Service d'Astrophysique, CEN Saclay, Gif-sur-Ivette, France
}                                                                              
\address{                                                                     
Naval Research Laboratory, Washington, D.C., USA
}                                                                              
\address{                                                                     
Durham University, Durham, U.K.
}                                                                              
\address{                                                                     
Royal Observatory, Edinburgh, U.K.
}  
\address{                                                                     
European Southern Observatory, Garching, Germany
} 
\address{                                                                     
Istituto di Radioastronomia del CNR, Bologna, Italy
}                                                                              
\begin{abstract}                                                                
Based on the ROSAT All-Sky Survey
we have conducted a large redshift survey as an ESO key programme
to identify and secure redshifts for the X-ray brightest clusters found
in the southern hemisphere. We present first results for a highly
controlled sample for a flux limit of $3\cdot 10^{-12}$ erg s$^{-1}$ 
cm$^{-2}$ (0.1 - 2.4 keV) comprising 475 clusters
(87\%  with redshifts). 
The logN-logS function of the sample shows an almost perfect Euclidian slope
and a preliminary X-ray luminosity function is presented.
\end{abstract}                                                                  
\section{Introduction}
For the study of the structure of the present day Universe on very 
large scales
($\ge 50 h^{-1}$ Mpc) the use of galaxy clusters 
constitutes a very interesting
alternative to the conventional galaxy redshift surveys for several
reasons. One can study a larger volume with a smaller number of objects.
The spatial correlation in the cluster distribution is strongly magnified
with respect to the galaxy and 
to the mass distribution (Kaiser 1984). The biasing
factor relating the cluster distribution power spectrum to the 
mass density fluctuations can be calculated {\it ab-initio}
(e.g. Bardeen et al. 1986, Mo \& White, 1996).

X-ray astronomy offers a unique tool to efficiently detect and             
characterize galaxy clusters out to large distances. Originating in the         
hot intracluster plasma that fills the gravitational potential well of          
the clusters, the X-ray emission is an equally robust parameter for a           
mass estimate of a clusters as the galaxy velocity
dispersion. With the survey described here, named ROSAT 
ESO Flux Limited X-ray (REFLEX) Cluster Survey, we exploit the 
unique opportunity provided by the ROSAT All-Sky Survey
(Tr\"umper 1993, Voges et al. 1996) to construct a
cluster sample for cosmological studies.
The cluster candidates found are then optically identified and
redshifts are measured  
in the frame of an ESO key programme (B\"ohringer 1994, Guzzo et al. 1995).
\begin{figure}                                                                  
\centerline{\vbox{                                                              
\psfig{figure=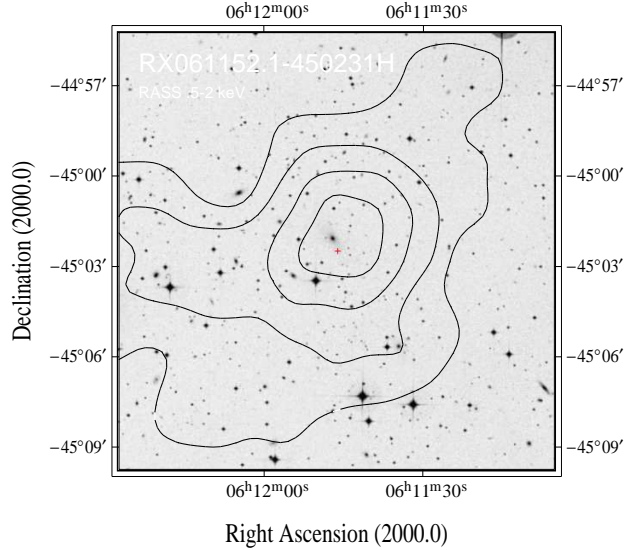,height=7.0cm}           
}}                                                                              
\caption[]{A typical, previously unknown cluster (with $z = 0.102$) 
from the REFLEX survey.
The digital sky map from the STScI scans of the UK Schmidt plates
is shown with X-ray surface brightness contours 
from the ROSAT All-Sky Survey superposed.}          
\end{figure}                                                                    
\section{Cluster Identification }
In the ROSAT All-Sky Survey atlas
only the brightest and well extended X-ray cluster sources are
readily identified, while the
main part of the identifications has to be based on further
optical information. For a first identification we use
the COSMOS data base (e.g. Heydon-Dumbleton et al. 1989), originating from
the UK Schmidt Survey, providing
star/galaxy separation 
down to $b_j \sim 20.5$ mag. 
The price paid for a high completeness (low detection threshold) 
is a contamination of the candidate list by 
more than 30\% non-cluster sources. This 
contamination is reduced by a direct inspection
of the photographic plates, the detailed X-ray properties, and
the available literature information. The residual contamination
(up to 10\%) is generally recognized and discarded in the follow-up 
observations, which will be completed at the end of the year 1998.
Presently we have constructed a first catalogue of bright clusters
down to an X-ray flux limit of
$3 \cdot 10^{-12}$ erg s$^{-1}$ cm$^{-2}$ (in the ROSAT band 0.1 - 2.4 keV)
comprising 475 objects.
Interestingly, only 53\% of these clusters are found in the  
ACO catalogue (Abell, Corwin, \& Olowin, 1989) and further 10\%
in the supplementary list,  while most of the others were
previously unknown (e.g. Fig. 1).
%
\begin{figure}                                                                  
\centerline{\vbox{                                                              
\psfig{figure=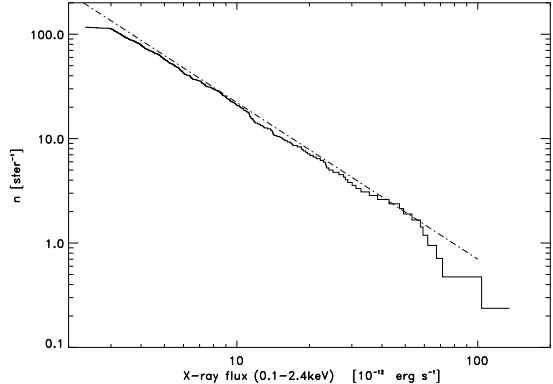,height=5.5cm}                                          
}}                                                                              
\caption[]{Number counts of clusters in the REFLEX sample as a function         
of the X-ray flux. The dashed line indicates a slope of -1.5.
}                                                                               
\end{figure}                                                                    
%
\section{Overall Properties of the REFLEX Clusters }
The following results are based on the sample of
475 X-ray bright clusters (with 413 cluster redshifts obtained
so far). A plot of the number counts of the population
as a function of X-ray flux is shown in Fig.\ 2. The logarithmic
graph is well described by an Euclidian slope of -3/2.
This is expected for such nearby clusters with a median redshift of
$z \sim 0.08$.

The X-ray luminosity function is a very important characteristic of the sample, 
since it is most closely related to the mass
function of the clusters and used as an important calibrator
of the amplitude of the cosmic density fluctuation power spectrum 
(e.g. White et al. 1993). A preliminary version of the REFLEX X-ray
luminosity function is shown in Fig. 3. The function was computed when
($ \sim 80 \%$) of the redshifts had been determined. But it 
already recovers the densities reached in previous surveys
(e.g. De Grandi 1996, Ebeling et al. 1997) as shown in Fig. 3.

\section{Conclusions}
Despite the present incompleteness in redshifts which we essentially hope
to fill by scheduled observations till the end of 1998, 
the high quality and completeness of the data set is already reflected in 
the present results. The large volume covered and the high accuracy
of the sample makes the REFLEX survey ideal for the study
of the large-scale structure. 
An extended REFLEX sample down to a flux limit of $2\cdot 10^{-12}$ erg
s$^{-1}$ cm$^{-2}$ is prepared and redshifts are available for
more than 70\% of the objects. This set will contain about 750 clusters. 
Finally, a complementary ROSAT Survey cluster identification
programme is being conducted in the Northern Sky in a collaboration of
MPE  and
J. Huchra, R. Giacconi, P. Rosati and B. McLean which will soon
reach a similar depth and provide an all-sky view on the X-ray cluster
distribution.

\begin{figure}                                                                  
\centerline{\vbox{                                                              
\psfig{figure=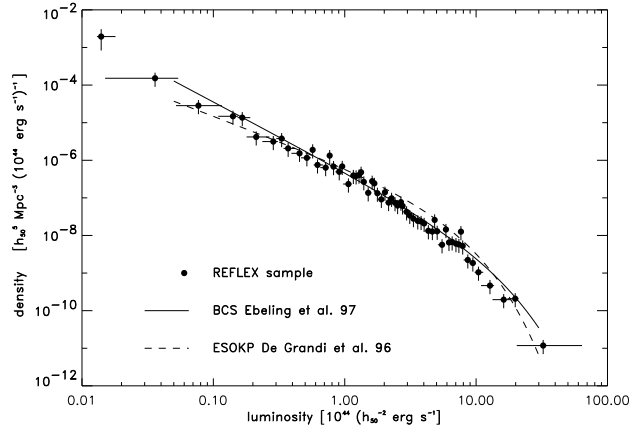,height=6.cm}                                              
}}                                                                              
\caption[]{X-ray luminosity function for 80\% of the clusters in the                    
REFLEX sample is compared to earlier ROSAT Survey studies by                                    
Ebeling et al. 1997 and De Grandi 1996.                                          
}                                                                               
\end{figure}                                                                    

\acknowledgements{ }      
We like to thank the ROSAT team and the COSMOS team at Edinburgh
and at NRL for help in the analysis of the X-ray and optical data.
                                                                                

%
\vfill                                                                          
\end{document}